\documentclass[runningheads]{llncs}
\usepackage{graphicx}
\usepackage{booktabs}
\usepackage{multirow}
\usepackage{graphicx}
\usepackage{pifont}
\usepackage{enumitem}
\newcommand{\cmark}{\ding{51}}%
\newcommand{\xmark}{\ding{55}}%
\newcommand{\etal}{\textit{et al.}}
\newcommand{\header}[1]{\vspace{1mm}\noindent\textbf{#1.}}

\begin{document}
\title{Improving the Robustness of Dense Retrievers Against Typos via Multi-Positive Contrastive~Learning}

\titlerunning{Typo-Robust Dense Retrieval via Multi-positive Contrastive Learning}

\author{Georgios Sidiropoulos\orcidID{0000-0002-6486-089X} \and
Evangelos Kanoulas\orcidID{0000-0002-8312-0694}
}
\institute{University of Amsterdam, Amsterdam, The Netherlands \\
\email\{g.sidiropoulos, e.kanoulas\}@uva.nl
}

\maketitle              
\begin{abstract}
Dense retrieval has become the new paradigm in passage retrieval. Despite its effectiveness on typo-free queries, it is not robust when dealing with queries that contain typos.
Current works on improving the typo-robustness of dense retrievers combine (i) data augmentation to obtain the typoed queries during training time with (ii) additional robustifying subtasks that aim to align the original, typo-free queries with their typoed variants.
Even though multiple typoed variants are available as positive samples per query, some methods assume a single positive sample and a set of negative ones per anchor and tackle the robustifying subtask with contrastive learning; therefore, making insufficient use of the multiple positives (typoed queries).
In contrast, in this work, we argue that all available positives can be used at the same time and employ contrastive learning that supports multiple positives (multi-positive).
Experimental results on two datasets show that our proposed approach of leveraging all positives simultaneously and employing multi-positive contrastive learning on the robustifying subtask yields improvements in robustness against using contrastive learning with a single positive.

\end{abstract}

\section{Introduction}
\label{sec:intro}
Dense retrieval has become the new paradigm in passage retrieval.
It has demonstrated higher effectiveness than traditional lexical-based methods due to its ability to tackle the vocabulary mismatch problem \cite{DBLP:conf/emnlp/KarpukhinOMLWEC20}.
Even though dense retrievers are highly effective on typo-free queries, they can witness a dramatic performance decrease when dealing with queries that contain typos \cite{DBLP:conf/sigir/SidiropoulosK22,DBLP:conf/cikm/SidiropoulosVK22,DBLP:conf/sigir/ZhuangZ22}. Recent works on robustifying dense retrievers against typos utilize data augmentation to obtain typoed versions of the original queries at training time. Moreover, they introduce additional robustifying subtasks to minimize the representation discrepancy between the original query and its typoed variants.

Sidiropoulos and Kanoulas \cite{DBLP:conf/sigir/SidiropoulosK22} applied an additional contrastive loss to enforce the latent representations of the original, typo-free queries to be closer to their typoed variants. 
Zhuang and Zuccon \cite{DBLP:conf/sigir/ZhuangZ22} utilized a self-teaching training strategy to minimize the difference between the score distribution of the original query and its typoed variants.
Alternatively, Tasawong \etal \cite{DBLP:conf/acl/TasawongPLUCN23} employed dual learning in combination with self-teaching \cite{DBLP:conf/sigir/ZhuangZ22} and contrastively trained the dense retriever on the prime task of passage retrieval and the dual task of query retrieval (learns the query likelihood to retrieve queries for passages). 

Despite the improvements in robustness, the existing typo-robust methods do not always make optimal use of the available typoed queries.
In detail, they address the robustifying subtasks with contrastive learning, assuming a single positive sample (query) and a set of negative ones per anchor (depending on the approach, the anchor can be either a query or a passage).
However, alongside the original query, its multiple typoed variants are available. Hence, there is more than one positive sample per anchor. As a result, we can leverage all the available positives simultaneously and apply multi-positive contrastive learning instead (i.e., contrastive learning that supports multiple positives).
For instance, Tasawong \etal \cite{DBLP:conf/acl/TasawongPLUCN23} computes the contrastive loss for the query retrieval subtask using only the original, typo-free query as relevant for a given passage. Given a passage, we argue that both the original query and its typoed variations can be considered as relevant and adopt a multi-positive contrastive loss instead.

Literature on contrastive learning has shown that including multiple positives can enhance the ability of the model to discriminate between signal and noise (negatives)\cite{DBLP:conf/nips/KhoslaTWSTIMLK20,9816043}.
Intuitively, multiple negatives focus on what makes the anchor and the negatives dissimilar, while multiple positives focus on what makes the anchor and the positives similar. 
To this end, contrasting among multiple positives and negatives can bring an anchor and all its positives closer together in the latent space while keeping them far from the negatives.

In this work, we revisit recent methods in typo-robust dense retrieval and unveil that, in many cases, they do not sufficiently utilize the multiple positives that are available.
Specifically, when tackling the robustifying subtasks, they ignore that multiple positives are available per anchor and consider contrastive learning with a single positive.
In contrast, we suggest leveraging all the available positives and adopting a multi-positive contrastive learning approach. We aim to answer the following research questions:
\begin{enumerate}[label=\textbf{RQ\arabic*},leftmargin=*]
% \textbf{RQ1}
\item Can our multi-positive contrastive learning approach increase the robustness of dense retrievers that use contrastive learning with a single positive?
% \textbf{RQ2} 
\item Does our multi-positive contrastive learning variant outperform its single-positive counterpart regardless of the number of positives?
\end{enumerate}
Our experimental results on two datasets show that our proposed approach of employing multi-positive contrastive learning yields improvements in robustness compared to contrastive learning with a single positive.
\footnote{\url{https://github.com/GSidiropoulos/typo-robust-multi-positive-DR}}

\section{Methodology}
\label{sec:method}
Contrastive learning is a vital component for training an effective dense retriever. Current typo-robust dense retrievers use contrastive learning with a single positive sample and multiple negative ones for both the main task of passage retrieval and the robustifying subtasks. In detail, given an anchor $x$, a positive sample $x^+$, and a set of negative samples $X^-$, the contrastive prediction task aims to bring the positive sample closer to the anchor than any other negative sample:
\begin{equation}
\mathcal{L}_{CE}(x,x^+,X^-) = -\log \frac{e^{f(x,x^+)}}{e^{f(x,x^+)} + \sum\limits_{x^- \in X^{-}}e^{f(x,x^-)}},
\label{eq:pce}
\end{equation}
where $f$ is a similarity function (e.g., dot product). 

However, in many cases, multiple positive samples are available per anchor and can be used simultaneously to increase the discriminative performance of the model. As opposed to the aforementioned contrastive loss that supports a single positive, we propose employing a multi-positive contrastive loss to benefit from all the available positives. Given an anchor $x$, multiple positive samples $X^+$, and multiple negatives $X^-$, a multi-positive contrastive loss \cite{DBLP:conf/nips/KhoslaTWSTIMLK20} is computed as:
\begin{equation}
\mathcal{L}_{MCE}(x,X^+,X^-) =  -\frac{1}{\vert X^+ \vert}\sum_{x^+ \in X^+} \log \frac{e^{f(x,x^+)}}{e^{f(x,x^+)} + \sum\limits_{x^- \in X^-}e^{f(x,x^-)}}.
\label{eq:mce}
\end{equation}

This work aims to identify cases in typo-robust dense retrieval methods where the robustifying subtasks consider only a single positive sample, even though multiple ones are available, and optimize a contrastive loss. Next, we replace the contrastive loss with its multi-positive alternative to benefit from all the available positives. Below we present the typo-robust dense retrieval methods we build upon followed by our multi-positive variants. We focus on dense retrievers that follow the dual-encoder architecture \cite{DBLP:conf/emnlp/KarpukhinOMLWEC20}. A traditional dense retriever, \textbf{DR}, is optimized only with the passage retrieval task. Given a query  $q$, a positive/relevant passage  $p^+$, and a set of negative/irrelevant passages $P^-=\{p_i^-\}_{i=1}^N$, the learning task trains the query and passage encoders via minimizing the softmax cross-entropy:
$\mathcal{L}^p_{CE} = \mathcal{L}_{CE}(q,p^+,P^-).
\label{eq:pce}
$
Positive query-passage pairs are encouraged to have higher similarity scores and negative pairs to have lower scores.

\subsection{Dense Retriever with Self-supervised Contrastive Learning}
\label{sec:dr_cl}
\textbf{DR+CL} alternates DR with an additional contrastive loss that maximizes the agreement between differently augmented views of the same query \cite{DBLP:conf/sigir/SidiropoulosK22}. This loss enforces that a query $q$ and its typoed variation $q'$, sampled from a set of available typoed variations $Q'=\{q_i'\}_{i=1}^K$, are close together in the latent space and distant from other distinct queries $Q^-=\{q_i^-\}_{i=1}^M$:
$
\mathcal{L}^t_{CE} = \mathcal{L}_{CE}(q,q',Q^-).
\label{eq:cl}
$
The final loss is computed as a weighted summation, $\mathcal{L}=w_1\mathcal{L}^p_{CE}+w_2\mathcal{L}^t_{CE}$.

\noindent \textbf{DR+CL$_M$} is our multi-positive variant of DR+CL. 
Given a query $q$, instead of sampling a different typoed variant $q'$ from a set $Q'$ at each update, we propose simultaneously employing all typoed variants.
To do so, we replace $\mathcal{L}^t_{CE}$ with the following multi-positive contrastive loss that accounts for multiple positives:
$
\mathcal{L}^t_{MCE} = \mathcal{L}_{MCE}(q,Q',Q^-).
\label{eq:qce}
$
The final loss is: $\mathcal{L}=w_1\mathcal{L}^p_{CE}+w_2\mathcal{L}^t_{MCE}$.

\subsection{Dense Retriever with Dual Learning}
\label{sec:dr_dl}
\textbf{DR+DL} trains a robust, dense retriever via a contrastive dual learning mechanism \cite{DBLP:conf/ictir/LiLX021}. In contrast to classic DR, which is optimized for passage retrieval only ($\mathcal{L}^p_{CE}$), DR+DL is optimized for the prime task of passage retrieval (i.e., learns to retrieve relevant passages for queries) and the dual task of query retrieval (i.e., learns to retrieve relevant queries for passages). Therefore, given a passage $p$, a positive query $q^+$, and a set of negative queries $Q^-=\{q_i^-\}_{i=1}^M$, it further minimizes the loss for the dual task:
$
\mathcal{L}^q_{CE} = \mathcal{L}_{CE}(p,q^+,Q^-).
\label{eq:qce}
$
The dual training loss is added to the prime training loss to conduct contrastive dual learning and train the dense retriever. Specifically, the final loss is computed as $\mathcal{L}=\mathcal{L}^p_{CE}+w\mathcal{L}^q_{CE}$, where $w$ is used to weight the dual task loss.

\noindent\textbf{DR+DL$_M$} is our multi-positive variant of DR+DL. 
% Given a passage $p$, DR+DL uses a contrastive loss, considering only the original query $q$ as the positive query for the dual task of query retrieval.
Contrary to DR+DL, we propose that for the query retrieval task, given a passage $p$, we can have a set of relevant queries consisting of the typo-free query and its typoed variants, $\mathcal{Q}=\{q^+, q'_1, q'_2, \dots, q'_K\}$. Thus, we replace the contrastive loss of $\mathcal{L}^q_{CE}$ with a multi-positive contrastive loss, which can account for multiple relevant queries at the same time. We define the multi-positive contrastive loss for the dual task as:
$\mathcal{L}^q_{MCE}=\mathcal{L}_{MCE}(p,Q,Q^-).
\label{eq:qce}
$
The final loss is computed as $\mathcal{L}=\mathcal{L}^p_{CE}+w\mathcal{L}^q_{MCE}$.

\subsection{Dense Retriever with Dual Learning and Self-Teaching}
\label{sec:dr_st_dl}
\textbf{DR+ST+DL} trains a dense retriever with dual learning and self-teaching \cite{DBLP:conf/acl/TasawongPLUCN23}. Similar to DR+DL, it minimizes the $\mathcal{L}^p_{CE}$ and $\mathcal{L}^q_{CE}$ for the main task of passage retrieval and the subtask of query retrieval, respectively. The additional self-teaching mechanism distills knowledge from a typo-free query $q$ into its typoed variants $Q'=\{q_i'\}_{i=1}^K$ by forcing the model to match score distributions of misspelled queries to the score distribution of the typo-free query for both the passage retrieval and query retrieval task. 
This is achieved by minimizing the KL-divergence losses:
(i) $\mathcal{L}^p_{KL} = \frac{1}{K}\sum_{k=1}^K\mathcal{L}_{KL}(s'^k_p, s_p)$, where $\{s'^1_p, s'^2_p, \dots, s'^K_p\}$ and $s_p$ is the score distribution in a passage-to-queries direction (passage retrieval) for the typoed queries and the typo-free query, respectively and
(ii) $\mathcal{L}^q_{KL}=\frac{1}{K}\sum_{k=1}^K\mathcal{L}_{KL}(s'^k_q, s_q)$, where $\{s'^1_q, s'^2_q, \dots, s'^K_q\}$ and $s_q$ is the score distribution in a query-to-passages direction (query retrieval) for the typoed queries and the typo-free query, respectively.
The final loss is computed as the weighted summation of the four losses, $\mathcal{L}=(1-\beta)((1-\gamma)\mathcal{L}^p_{CE}+\gamma\mathcal{L}^q_{CE})+ \beta((1-\sigma)\mathcal{L}^p_{KL}+\sigma\mathcal{L}^q_{KL})$.

\noindent\textbf{DR+ST+DL$_M$} is our multi-positive variant of DR+ST+DL. Even though DR+ST+DL simultaneously uses all the available typo variations of a query in order to calculate the KL divergence losses for the prime passage retrieval task and the dual query retrieval, it uses only the typo-free query to compute the contrastive loss for query retrieval ($\mathcal{L}^q_{CE}$). To fully benefit from the multiple available typoed queries per typo-free query, we replace the contrastive loss for query retrieval $\mathcal{L}^q_{CE}$ with a multi-positive variant that supports samples with multiple positives $\mathcal{L}^q_{MCE}$. The final loss is computed as the weighted summation, $\mathcal{L}=(1-\beta)((1-\gamma)\mathcal{L}^p_{CE}+\gamma\mathcal{L}^q_{MCE})+ \beta((1-\sigma)\mathcal{L}^p_{KL}+\sigma\mathcal{L}^q_{KL})$.

\section{Experimental Setup}
\label{sec:exp_setup}

\header{Query augmentation}
\label{sec:query_augm}
From the aforementioned methods, those employing queries with typos in their training scheme are augmentation-based. In detail, during training, the typoed queries are generated from the original, typo-free queries through a realistic typo generator \cite{DBLP:conf/emnlp/MorrisLYGJQ20}. The typo generator applies the following transformations that often occur in human-generated queries: random character insertion, deletion and substitution, swapping neighboring characters,  and keyboard-based character swapping \cite{DBLP:conf/sigir/HagenPGRS17}.

\header{Datasets and evaluation}
We conduct our experiments on MS MARCO passage ranking \cite{DBLP:conf/nips/NguyenRSGTMD16} and DL-Typo \cite{DBLP:conf/sigir/ZhuangZ22} on their typo-free and typoed versions. Both datasets use the same underlying corpus of $8.8$ million passages and $\sim400K$ training queries but differ in evaluation queries.
DL-Typo provides $60$ real-world queries with typos alongside their manually corrected typo-free version.
The development set of MS MARCO consists of $6,980$ queries (the test set is not publicly available). Following previous works \cite{DBLP:conf/acl/TasawongPLUCN23,DBLP:conf/sigir/ZhuangZ22}, we obtain typo variations for each typo-free query via a synthetic typo generation model and repeat the typo generation process $10$ times.
To measure the retrieval performance, we report the official metrics of each dataset. For the evaluation on the typo version of MS MARCO, we report the metrics averaged for each repeated experiment since typoed queries are generated $10$ times for each original query.

\header{Implementation details}
We follow an in-batch negative training setting with $7$ hard negative passages per query and a batch size of $16$ to train the dense retrievers.
\footnote{The original methods and our proposed counterparts employ the same number of original, typo-free query-passage pairs per batch. However, our method leverages multiple typoed variants for each query; therefore, the batch we need to fit in the GPU memory is larger.}
We use AdamW optimizer with a $10^{-5}$ learning rate, linear scheduling with $10K$ warm-up steps, and decay over the rest of the training steps. We train up to $150K$ steps. We implement the query and passage encoders with BERT \cite{DBLP:conf/naacl/DevlinCLT19}.
When applicable, we set the query augmentation size to 40.
For the remaining hyperparameters specific to each method (e.g., weight $w$ in DR+CL), we use the values initially proposed by the creators of each method.
We use the Tevatron toolkit \cite{DBLP:conf/sigir/GaoMLC23} to train the models and the Ranx library \cite{DBLP:conf/ecir/Bassani22} to evaluate the retrieval performance.
Finally, we use the typo generators from the TextAttack toolkit \cite{DBLP:conf/emnlp/MorrisLYGJQ20} for all the methods we experiment with to augment the training queries.

\section{Results}
\label{sec:results}
To answer \textbf{RQ1}, we compare the retrieval performance of our multi-positive contrastive learning approaches against the original models. From Table \ref{tab:multi_vs_single}, we see that employing our multi-positive contrastive learning approach yields improvements in robustness against typos upon the original methods that use contrastive learning with a single positive. 

As expected, the more dramatic improvement comes when applying multi-positive contrastive learning on DR+DL since the original work only considers the typo-free query as positive when computing the contrastive loss for query retrieval (see Section \ref{sec:dr_dl}). In contrast, in our DR+DL$_M$, we consider the typo-free query and all its available typoed variants as positives and use a multi-positive contrastive loss for query retrieval. We also see improvements when comparing DR+CL vs. our DR+CL$_M$. In detail, employing all available positives (typoed queries) at once and using multi-positive contrastive loss outperforms sampling a different positive at each update and using a single positive contrastive loss (see Section \ref{sec:dr_cl}). The improvements are held even when comparing our DR+DL+ST$_M$ against DR+DL+ST, a model that already uses multiple positives. As seen in Section \ref{sec:dr_st_dl}, DR+DL+ST uses a contrastive loss with a single positive for the query retrieval dual task (i.e., $\mathcal{L}^q_{CE}$) while considering multiple positives simultaneously to compute the KL-divergence losses (i.e., $\mathcal{L}^p_{KL}$, $\mathcal{L}^q_{KL}$).

\begin{table}[ht!]
\caption{Retrieval results for the settings of (i) clean queries (Clean), and (ii) queries with typos (Typo). Statistical significant gains (two-tailed paired t-test with Bonferroni correction, $p<0.05$) obtained from models with multi-positive contrastive loss (ours) over their original version with standard contrastive loss are indicated by $\dagger$.}
\tabcolsep=0.12cm
\resizebox{\columnwidth}{!}{%
\begin{tabular}{@{}lccccc|cllcll@{}}
\hline
\multirow{3}{*}{Model} & \multirow{3}{*}{\begin{tabular}[c]{@{}l@{}}Multi-positive\\ contrastive loss\end{tabular}} & \multicolumn{4}{c|}{MS MARCO} & \multicolumn{6}{c}{DL-Typo} \\
 &  & \multicolumn{2}{c}{Clean} & \multicolumn{2}{c|}{Typo} & \multicolumn{3}{c}{Clean} & \multicolumn{3}{c}{Typo} \\
 &  & MRR@10 & R@1000 & MRR@10 & R@1000 & nDCG@10 & MRR & MAP & nDCG@10 & MRR & MAP \\ \hline
DR & \xmark & .331 & .953 & .140 & .698 & .677 & .850 & .555 & .264 & .395 & .180 \\ \hline
DR+DL & \xmark & .332 & .953 & .140 & .698 & .679 & .826 & .557 & .269 & .411 & .186 \\
DR+DL$_M$ & \cmark & .335 & .958 & .213$^\dagger$ & .866$^\dagger$ & .699 & .864 & .585 & .347$^\dagger$ & .452 & .259$^\dagger$ \\ \hline
DR+CL & \xmark & .321 & .957 & .170 & .787 & .659 & .797 & .535 & .284 & .411 & .207 \\
DR+CL$_M$ & \cmark & .322 & .956 & .178$^\dagger$ & .811$^\dagger$ & .652 & .847 & .539 & .290 & .447 & .215 \\ \hline
DR+ST+DL & \xmark & .334 & .951 & .259 & .893 & .681 & .868 & .567 & .412 & .543 & .315 \\
DR+ST+DL$_M$ & \cmark & .335 & .955 & .261 & .902$^\dagger$ & .687 & .870 & .579 & .426$^\dagger$ & .583 & .342$^\dagger$ \\ \hline
\end{tabular}%
}
\label{tab:multi_vs_single}
\end{table}

At this point, we want to explore how the different numbers of positives affect our multi-positive approach (\textbf{RQ2}). To do so, we compare our DR+DL+ST$_M$ against DR+DL+ST. In its training, the latter already employs multiple positives simultaneously to compute the KL-divergence losses. However, our multi-positive approach fully benefits from the multiple available positives by incorporating them when computing the contrastive loss for query retrieval ($\mathcal{L}^q_{CE}\rightarrow\mathcal{L}^q_{MCE}$). Table \ref{tab:diff_K} unveils that our multi-positive variant consistently outperforms the original model for the different numbers of typoed variants per query.

\begin{table}[ht!]
\caption{Retrieval results for different query augmentation sizes ($K$). We report the results
in the format ``$R@1000\ (MRR@10)$'' on MS MARCO with typos.}
\resizebox{\columnwidth}{!}{%
\tabcolsep=0.12cm
\begin{tabular}{@{}l c l l l l l @{}}
\toprule
         & \multirow{2}{*}{\begin{tabular}[c]{@{}l@{}}Multi-positive\\ contrastive loss\end{tabular}}           & \multicolumn{4}{c}{$K$}                                 &             \\
         & & 1           & 10          & 20          & 30          & 40          \\ \midrule
DR+ST+DL &   \xmark          & .884 (.251) & .892 (.258) & .894 (.258) & .893 (.259) & .893 (.259) \\
DR+ST+DL$_M$ &   \cmark          & \textbf{.884 (.251)} & \textbf{.898 (.260)} & \textbf{.900 (.260)} & \textbf{.902 (.261)} & \textbf{.902 (.261)} \\ \bottomrule
\end{tabular}
}
\label{tab:diff_K}
\end{table}

\section{Conclusions}
\label{sec:conclusions}

In this work, we revisit recent studies in typo-robust dense retrieval and showcase that they do not always make sufficient use of multiple positive samples. In detail, they assume a single positive sample and multiple negatives per anchor and use contrastive learning for the robustifying subtasks. Opposed to this, we propose to leverage all the available positives and employ multi-positive contrastive learning. Experimentation on two datasets shows that following a multi-positive contrastive learning approach yields improvements in the robustness of the underlying dense retriever upon contrastive learning with a single positive.

\subsubsection{Acknowledgements} 
This research was supported by
the NWO Innovational Research Incentives Scheme Vidi (016.Vidi.189.039).
% 
% the NWO Smart Culture - Big Data / Digital Humanities (314-99-301),
%
% the H2020-EU.3.4. - SOCIETAL CHALLENGES - Smart, Green And Integrated Transport (814961).
%
All content represents the opinion of the authors, which is not necessarily shared or endorsed by their respective employers and/or sponsors.

\bibliographystyle{splncs04}
\bibliography{mybibliography}

\begin{thebibliography}{10}
\providecommand{\url}[1]{\texttt{#1}}
\providecommand{\urlprefix}{URL }
\providecommand{\doi}[1]{https://doi.org/#1}

\bibitem{DBLP:conf/ecir/Bassani22}
Bassani, E.: ranx: {A} blazing-fast python library for ranking evaluation and comparison. In: Hagen, M., Verberne, S., Macdonald, C., Seifert, C., Balog, K., N{\o}rv{\aa}g, K., Setty, V. (eds.) Advances in Information Retrieval - 44th European Conference on {IR} Research, {ECIR} 2022, Stavanger, Norway, April 10-14, 2022, Proceedings, Part {II}. Lecture Notes in Computer Science, vol. 13186, pp. 259--264. Springer (2022). \doi{10.1007/978-3-030-99739-7\_30}, \url{https://doi.org/10.1007/978-3-030-99739-7\_30}

\bibitem{DBLP:conf/naacl/DevlinCLT19}
Devlin, J., Chang, M., Lee, K., Toutanova, K.: {BERT:} pre-training of deep bidirectional transformers for language understanding. In: Burstein, J., Doran, C., Solorio, T. (eds.) Proceedings of the 2019 Conference of the North American Chapter of the Association for Computational Linguistics: Human Language Technologies, {NAACL-HLT} 2019, Minneapolis, MN, USA, June 2-7, 2019, Volume 1 (Long and Short Papers). pp. 4171--4186. Association for Computational Linguistics (2019). \doi{10.18653/v1/n19-1423}, \url{https://doi.org/10.18653/v1/n19-1423}

\bibitem{DBLP:conf/sigir/GaoMLC23}
Gao, L., Ma, X., Lin, J., Callan, J.: Tevatron: An efficient and flexible toolkit for neural retrieval. In: Chen, H., Duh, W.E., Huang, H., Kato, M.P., Mothe, J., Poblete, B. (eds.) Proceedings of the 46th International {ACM} {SIGIR} Conference on Research and Development in Information Retrieval, {SIGIR} 2023, Taipei, Taiwan, July 23-27, 2023. pp. 3120--3124. {ACM} (2023). \doi{10.1145/3539618.3591805}, \url{https://doi.org/10.1145/3539618.3591805}

\bibitem{DBLP:conf/sigir/HagenPGRS17}
Hagen, M., Potthast, M., Gohsen, M., Rathgeber, A., Stein, B.: A large-scale query spelling correction corpus. In: Kando, N., Sakai, T., Joho, H., Li, H., de~Vries, A.P., White, R.W. (eds.) Proceedings of the 40th International {ACM} {SIGIR} Conference on Research and Development in Information Retrieval, Shinjuku, Tokyo, Japan, August 7-11, 2017. pp. 1261--1264. {ACM} (2017). \doi{10.1145/3077136.3080749}, \url{https://doi.org/10.1145/3077136.3080749}

\bibitem{DBLP:conf/emnlp/KarpukhinOMLWEC20}
Karpukhin, V., Oguz, B., Min, S., Lewis, P.S.H., Wu, L., Edunov, S., Chen, D., Yih, W.: Dense passage retrieval for open-domain question answering. In: Webber, B., Cohn, T., He, Y., Liu, Y. (eds.) Proceedings of the 2020 Conference on Empirical Methods in Natural Language Processing, {EMNLP} 2020, Online, November 16-20, 2020. pp. 6769--6781. Association for Computational Linguistics (2020). \doi{10.18653/v1/2020.emnlp-main.550}, \url{https://doi.org/10.18653/v1/2020.emnlp-main.550}

\bibitem{DBLP:conf/nips/KhoslaTWSTIMLK20}
Khosla, P., Teterwak, P., Wang, C., Sarna, A., Tian, Y., Isola, P., Maschinot, A., Liu, C., Krishnan, D.: Supervised contrastive learning. In: Larochelle, H., Ranzato, M., Hadsell, R., Balcan, M., Lin, H. (eds.) Advances in Neural Information Processing Systems 33: Annual Conference on Neural Information Processing Systems 2020, NeurIPS 2020, December 6-12, 2020, virtual (2020), \url{https://proceedings.neurips.cc/paper/2020/hash/d89a66c7c80a29b1bdbab0f2a1a94af8-Abstract.html}

\bibitem{DBLP:conf/ictir/LiLX021}
Li, Y., Liu, Z., Xiong, C., Liu, Z.: More robust dense retrieval with contrastive dual learning. In: Hasibi, F., Fang, Y., Aizawa, A. (eds.) {ICTIR} '21: The 2021 {ACM} {SIGIR} International Conference on the Theory of Information Retrieval, Virtual Event, Canada, July 11, 2021. pp. 287--296. {ACM} (2021). \doi{10.1145/3471158.3472245}, \url{https://doi.org/10.1145/3471158.3472245}

\bibitem{9816043}
Małkiński, M., Mańdziuk, J.: Multi-label contrastive learning for abstract visual reasoning. IEEE Transactions on Neural Networks and Learning Systems pp. 1--13 (2022). \doi{10.1109/TNNLS.2022.3185949}

\bibitem{DBLP:conf/emnlp/MorrisLYGJQ20}
Morris, J.X., Lifland, E., Yoo, J.Y., Grigsby, J., Jin, D., Qi, Y.: Textattack: {A} framework for adversarial attacks, data augmentation, and adversarial training in {NLP}. In: Liu, Q., Schlangen, D. (eds.) Proceedings of the 2020 Conference on Empirical Methods in Natural Language Processing: System Demonstrations, {EMNLP} 2020 - Demos, Online, November 16-20, 2020. pp. 119--126. Association for Computational Linguistics (2020). \doi{10.18653/v1/2020.emnlp-demos.16}, \url{https://doi.org/10.18653/v1/2020.emnlp-demos.16}

\bibitem{DBLP:conf/nips/NguyenRSGTMD16}
Nguyen, T., Rosenberg, M., Song, X., Gao, J., Tiwary, S., Majumder, R., Deng, L.: {MS} {MARCO:} {A} human generated machine reading comprehension dataset. In: Besold, T.R., Bordes, A., d'Avila Garcez, A.S., Wayne, G. (eds.) Proceedings of the Workshop on Cognitive Computation: Integrating neural and symbolic approaches 2016 co-located with the 30th Annual Conference on Neural Information Processing Systems {(NIPS} 2016), Barcelona, Spain, December 9, 2016. {CEUR} Workshop Proceedings, vol.~1773. CEUR-WS.org (2016), \url{http://ceur-ws.org/Vol-1773/CoCoNIPS\_2016\_paper9.pdf}

\bibitem{DBLP:conf/sigir/SidiropoulosK22}
Sidiropoulos, G., Kanoulas, E.: Analysing the robustness of dual encoders for dense retrieval against misspellings. In: Amig{\'{o}}, E., Castells, P., Gonzalo, J., Carterette, B., Culpepper, J.S., Kazai, G. (eds.) {SIGIR} '22: The 45th International {ACM} {SIGIR} Conference on Research and Development in Information Retrieval, Madrid, Spain, July 11 - 15, 2022. pp. 2132--2136. {ACM} (2022). \doi{10.1145/3477495.3531818}, \url{https://doi.org/10.1145/3477495.3531818}

\bibitem{DBLP:conf/cikm/SidiropoulosVK22}
Sidiropoulos, G., Vakulenko, S., Kanoulas, E.: On the impact of speech recognition errors in passage retrieval for spoken question answering. In: Hasan, M.A., Xiong, L. (eds.) Proceedings of the 31st {ACM} International Conference on Information {\&} Knowledge Management, Atlanta, GA, USA, October 17-21, 2022. pp. 4485--4489. {ACM} (2022). \doi{10.1145/3511808.3557662}, \url{https://doi.org/10.1145/3511808.3557662}

\bibitem{DBLP:conf/acl/TasawongPLUCN23}
Tasawong, P., Ponwitayarat, W., Limkonchotiwat, P., Udomcharoenchaikit, C., Chuangsuwanich, E., Nutanong, S.: Typo-robust representation learning for dense retrieval. In: Rogers, A., Boyd{-}Graber, J.L., Okazaki, N. (eds.) Proceedings of the 61st Annual Meeting of the Association for Computational Linguistics (Volume 2: Short Papers), {ACL} 2023, Toronto, Canada, July 9-14, 2023. pp. 1106--1115. Association for Computational Linguistics (2023). \doi{10.18653/v1/2023.acl-short.95}, \url{https://doi.org/10.18653/v1/2023.acl-short.95}

\bibitem{DBLP:conf/sigir/ZhuangZ22}
Zhuang, S., Zuccon, G.: Characterbert and self-teaching for improving the robustness of dense retrievers on queries with typos. In: Amig{\'{o}}, E., Castells, P., Gonzalo, J., Carterette, B., Culpepper, J.S., Kazai, G. (eds.) {SIGIR} '22: The 45th International {ACM} {SIGIR} Conference on Research and Development in Information Retrieval, Madrid, Spain, July 11 - 15, 2022. pp. 1444--1454. {ACM} (2022). \doi{10.1145/3477495.3531951}, \url{https://doi.org/10.1145/3477495.3531951}

\end{thebibliography}

\end{document}